\documentclass{cmslatex}
\usepackage{latexsym, amssymb, enumerate, amsmath}

\usepackage{graphicx}
\usepackage{lscape}

\newcommand{\EE}{\mathbb{E} }

\newcommand{\PP}{\mathbb{P} }

\def\o{\overline}

\usepackage{color}

\newcommand{\be}{\begin{equation}}
\newcommand{\en}{\end{equation}}

\renewcommand{\appendixname}{Appendix}

\newcommand{\ea}{\end{eqnarray}}
\newcommand{\ba}{\begin{eqnarray}}
\newcommand{\ean}{\end{eqnarray*}}
\newcommand{\ban}{\begin{eqnarray*}}

\begin{document}

\title{Mean Field Games and Systemic Risk}

\author{Ren\'e Carmona\thanks{ORFE, Bendheim Center for Finance, Princeton University, Princeton, NJ 08544, {\em
rcarmona@princeton.edu}. Partially supported by NSF grant DMS-0806591.}
\and Jean-Pierre Fouque\thanks{Department of Statistics \& Applied Probability,
 University of California,
        Santa Barbara, CA 93106-3110, {\em fouque@pstat.ucsb.edu}. Work  supported by NSF grant DMS-1107468.}
        \and Li-Hsien Sun\thanks{Department of Statistics \& Applied Probability,
 University of California,
        Santa Barbara, CA 93106-3110, {\em sun@pstat.ucsb.edu}.}}
\date{\today}
\pagestyle{myheadings} \markboth{Mean Field Games and Systemic Risk}{R. Carmona, J.-P Fouque,   L.-H. Sun}\maketitle

\bigskip

\begin{abstract}

We propose a simple model of inter-bank borrowing and lending where the evolution of the log-monetary reserves of $N$
banks is described by a system of diffusion processes coupled through their drifts in such a way that stability of the
system depends on the rate of inter-bank borrowing and lending. Systemic risk is characterized
by a large number of banks reaching a default threshold by a given time horizon. 
 Our model incorporates a  game feature  where each bank
controls its rate of borrowing/lending to a central bank. The optimization reflects the desire of each
bank to borrow from the central bank when its monetary reserve falls below a critical level or lend if it rises above
this critical level which is chosen here as the average monetary reserve. Borrowing from or lending to the central
bank is also subject to a quadratic cost at a rate which can be fixed by the regulator. We solve explicitly for Nash equilibria with finitely many players, and we show that in this model the central bank acts as a clearing house, adding liquidity to the system without affecting its systemic risk. We also study the corresponding Mean Field Game in the limit of large number of banks in the presence of a common noise.

\end{abstract}


\centerline{\it Dedicated to George Papanicolaou in honor of his 70th birthday}

\section{Introduction}

 Systemic risk is becoming a central research topic. We refer to the  Handbook
\cite{handbook} for recent developments on systemic risk from many points of view (Statistics, Finance, Mathematical
Finance, Behavioral Finance, Networks, Counterparty Risk, High Frequency Trading, ...). Here,
we propose a simple model of inter-bank borrowing and lending where the evolution of the log-monetary reserves of $N$
banks is described by a system of diffusion processes coupled through their drifts in such a way that stability of the
system depends on the rate of inter-bank borrowing and lending. Systemic risk is characterized
by a large number of banks reaching a default threshold by a given time horizon. This type of interaction and the
relation {\it stability--systemic risk} has been recently studied in  \cite{Fouque-Ichiba}, \cite{Fouque-Sun},
\cite{Garnier-Mean-Field}, and \cite{GarnierPapanicolaouYang}. Here, we  introduce a game feature  where each bank
controls its rate of borrowing/lending to a central bank. The control of each individual bank reflects the desire to borrow from the central bank when its monetary reserve falls below a critical level or lend if it rises above
this critical level which is chosen here as the average monetary reserve. Borrowing from or lending to the central
bank is also subject to a quadratic cost at a rate which can be fixed by the regulator. As written, our
model is an example of Linear-Quadratic Mean Field Game with finitely many players which can be solved explicitly. We first solve for open-loop equilibria using the Pontryagin stochastic maximum principle. We also solve for closed-loop equilibria using the probabilistic approach based on the Pontryagin stochastic maximum principle leading to the solution of  Forward-Backward Stochastic Differential Equations, and the dynamic programming principle leading to the solution of Hamilton-Jacobi-Bellman partial differential equations. We also study the corresponding Mean Field Game in the limit of a large number of banks. Adding a game component to the model
has a non-trivial effect on the stability of the system which our simple model enables  to analyze. Our conclusion is
that  inter-bank borrowing and lending  creates stability, this stability is enhanced by the possibility of borrowing and lending from a central bank which in our model appears as a clearing house providing additional liquidity, and systemic risk is described  as a rare event with probability quantified by large deviation theory.

In the  model discussed below, the diffusion processes $X^{i}_t$ for  $i=1,\dots, N$ represent the log-monetary
reserves of $N$ banks lending to and borrowing from each other. The system is driven by $N$ (possibly correlated) standard
Brownian motions $\widetilde{W}^i_t, i=1,\cdots,N$ written as $\widetilde{W}^i_t=\rho W^0_t+\sqrt{1-\rho^2}W^i_t$ where $W^j_t, j=0,1,\cdots,N$ are independent standard Brownian motions, $W^0_t$ being the {\it common noise}, and $|\rho|\leq 1$. The system starts at time $t=0$ from i.i.d. random variables $X^i_0=\xi^i$ independent of the Brownian motions and such that $\EE(\xi^i)=0$.
We assume that the diffusion coefficients are constant and identical, denoted by $\sigma >0$.
Our  model of lending and borrowing consists in introducing an interaction through drift terms representing the rate at which bank $i$ borrows from or lends to bank $j$. In this case, the
rates are proportional to the difference in log-monetary reserves, and our model is:
\be\label{coupled}
d X^{i}_t =\frac{a}{N}\sum_{j=1}^N(X^{j}_t-X^{i}_t)\,dt
+\alpha^i_t dt
 +\sigma d \widetilde{W}^{i}_t\,, \quad  i=1,\dots, N,
\en
where the overall rate of ``mean-reversion" $a/N$ has been normalized by the number of banks with  $a\geq 0$. Bank
$i$ controls its rate of borrowing/lending to a central bank through the control rate $\alpha^i_t $.
Using the notation
$$\overline{X}_t=\frac{1}{N}\sum_{i=1}^NX^i_t,$$ 
for the empirical mean, the dynamics can be rewritten in the mean field form:
\be\label{coupled2}
dX^i_t=\left[a(\overline{X}_t-X^i_t)+{\alpha^i_t}\right]dt +\sigma d\widetilde{W}^i_t, \quad i=1,\cdots,N.
\en
Bank $i\in\{1,\cdots,N\}$ controls its rate of lending and borrowing at time $t$ by choosing the control $\alpha^i_t$ in order to minimize
\be\label{cost}
J^i(\alpha^1,\cdots,\alpha^N)=\EE\left\{\int_0^T f_i(X_t,\alpha^i_t)dt+g_i(X_{T}^{i})\right\},
\en
where the running cost function $f_i$ is defined by
\begin{equation}
\label{fo:f}
f_i(x,\alpha^i)=\left[\frac{1}{2}(\alpha^i)^2-q\alpha^i(\overline{x}-x^i)+\frac{\epsilon}{2}(\overline{x}-x^i)^2\right],
\end{equation}
and the terminal cost function $g_i$ by
\begin{equation}
\label{fo:g}
g_i(x)=\frac{c}{2}\left(\overline{x}-x^{i}\right)^2.
\end{equation}
Notice that the running quadratic cost $\frac{1}{2}(\alpha^i)^2$ has been normalized and that the effect of the parameter  $q>0$ is to control the {\it incentive} to borrowing or lending:  
the bank $i$ will want to borrow ($\alpha^i_t>0$)  if $X^i_t$ is smaller than the empirical mean ($\overline{X}_t$) and lend ($\alpha^i_t<0$)
if $X^i_t$ is larger than $\overline{X}_t$. Equivalently, after dividing by $q>0$, this parameter can be thought as a control by the regulator of the cost of borrowing or lending (with $q$ large meaning low fees).

The quadratic terms in $\left(\overline{x}-x^{i}\right)^2$ in the running cost ($\epsilon>0$) and in the terminal cost ($c>0$) penalize departure from the average. We assume that
\ba\label{convexitycondition}
 q^2\leq\epsilon,
\ea
so that $f_i(x,\alpha)$ is convex in $(x,\alpha)$.

In the spirit of structural models of defaults, we introduce a default level $D <0$ and say that bank $i$ defaults by
time $T$ if its log-monetary reserve reached the level $D$ before time $T$. Note that in this simple model, even after
reaching the default level, bank $i$ stays in the system until time $T$ and continues to participate in inter-bank and
central bank borrowing and lending activities.

The paper is organized as follows.
In Section \ref{sec:nogame} we recall the results of the descriptive model presented in \cite{Fouque-Sun} without control,  that is  $\alpha^i_t=0$. We illustrate the fact that inter-bank borrowing and lending creates {\it stability}, and  we define and quantify the {\it systemic risk}.

Section \ref{sec:game} is devoted to the analysis of the stochastic differential game  (\ref{coupled2}-\ref{cost}). We derive exact Nash equilibria for the open-loop as well as for the closed-loop Markovian models, using both probabilist and analytic approaches. 

The financial implications in terms of liquidity and role of a central bank are discussed in Section \ref{sec:financial}.

Stochastic differential games with a large number of players are usually not tractable. It is because of the very special nature of our model (linear dynamics, quadratic costs, and interactions through the empirical mean) that we are able to construct explicitly Nash equilibria, both open and closed loops. For generic models which are not amenable to explicit solutions, Lasry and Lions \cite{MFG1,MFG2,MFG3} have recently provided an elegant way to tackle the construction of approximate Nash equilibria for large games with mean field interactions.
Their methodology, known as  {\it Mean Field Game} (MFG), has been applied to a wide variety of problems (see \cite{GueantLasryLions.pplnm,Lachapelle} for some examples). A similar research program was developed independently by Caines, Huang, and Malham\'{e} with the name of Nash Certainty Equivalent. See for example \cite{HuangCainesMalhame1} and \cite{HuangCainesMalhame2}. 
The approach of Lasry and Lions (e.g. \cite{MFG3})  is based on the solution of a system of partial differential equations (PDEs): a Hamilton-Jacobi-Bellman equation evolving backward in time, and a Kolmogorov equation evolving forward in time, these two PDEs being strongly coupled. By its probabilistic nature, it is natural to recast the MFG strategy using appropriate forms of the Pontryagin stochastic maximum principle, leading to the solution of new models. See, for example, \cite{CarmonaDelarueLachapelle,Bensoussan_et_al,CarmonaDelarue_sicon,CarmonaDelarue_ap} or \cite{CarmonaLacker1} for a probabilistic approach based on the weak formulation of stochastic control. We consider the MFG problem and discuss the existence of approximate Nash equilibria in Section \ref{sec:approximate}.

\section{Stability and Systemic Risk}\label{sec:nogame}

In this section, we consider a system of $N$ banks without the possibility of borrowing or lending to a central bank,
that is (\ref{coupled}) or (\ref{coupled2}) with $\alpha^i_t =0$:
\ba\label{couplednogame}
d X^{i}_t &=&\frac{a}{N}\sum_{j=1}^N(X^{j}_t-X^{i}_t)\,dt\nonumber
 +\sigma d \widetilde{W}^{i}_t\\
 &=&a(\overline{X}_t-X^i_t)dt +\sigma d\widetilde{W}^i_t, \quad i=1,\cdots,N.
\ea
To start with, we assume that the Brownian motions $\widetilde{W}^i$ are independent, that is $\rho=0$ and $\widetilde{W}^i=W^i$. For simplicity, we also assume that $X_0^i=\xi^i=0$. In Section \ref{sec:commonnoise}, we will consider the correlated (or {\it common noise}) case.
\subsection{Simulations}

\begin{figure}
\includegraphics[width=13cm,height=6cm]{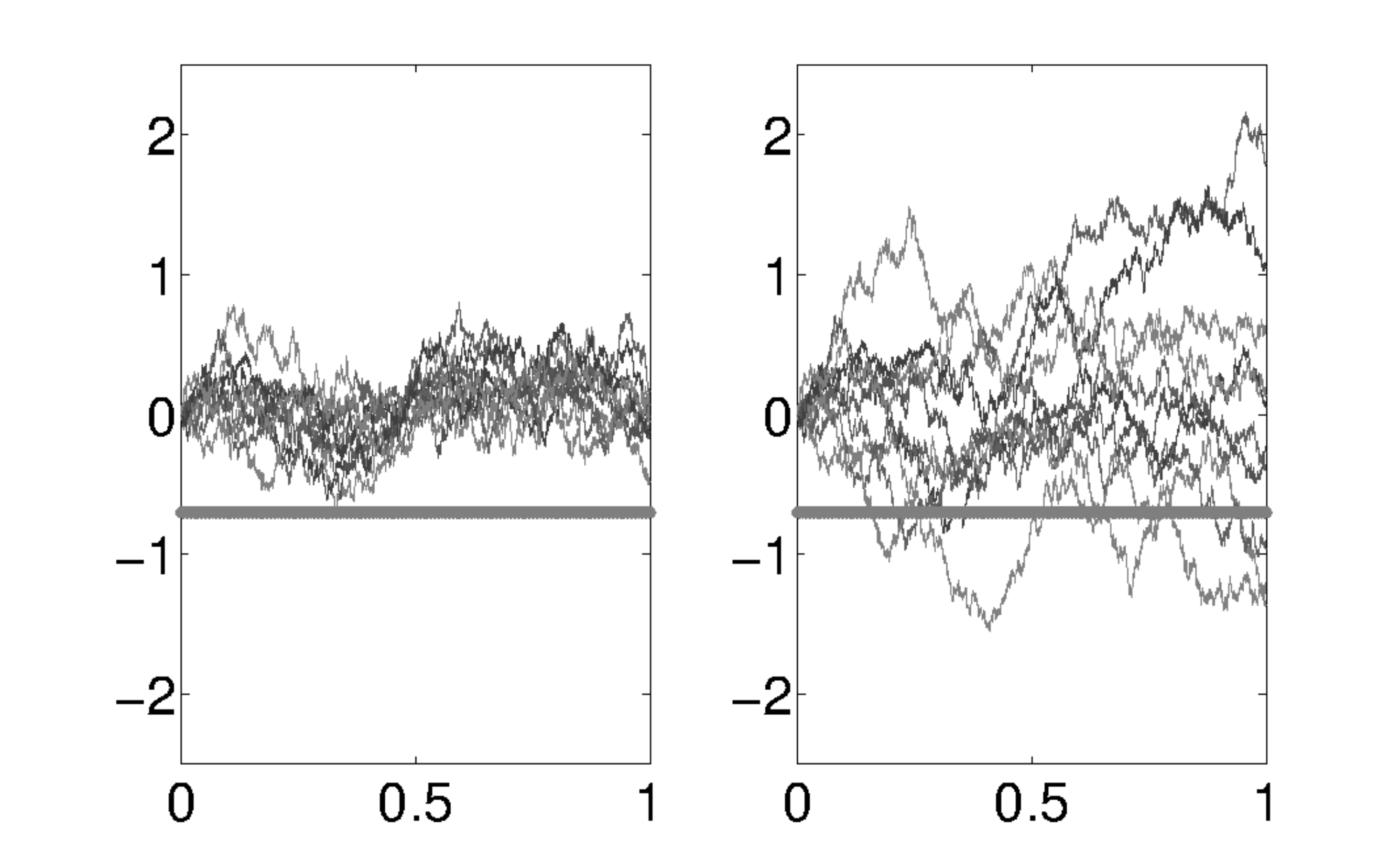}
\caption{One realization of  $N=10$ trajectories of the coupled diffusions (\ref{couplednogame}) (left plot) with $a=10$
and trajectories of  independent Brownian motions ($a=0$) (right plot) using the same Gaussian increments. The solid
horizontal line represents the ``default" level $D=-0.7$.}
\label{large alpha}
\end{figure}

In Figure  \ref{large alpha}  we show a typical realization of the $N$ trajectories with a relatively large rate of
borrowing/lending  $a=10$, the diffusion coefficient being set at $\sigma=1$ (we used the Euler scheme with a
time-step $\Delta=10^{-4}$, up to time $T=1$).
We see that the trajectories generated by (\ref{couplednogame}) are more grouped than the ones generated by
independent Brownian motions corresponding to no inter-bank borrowing or lending ($a=0$). This is the ``swarming" or
``flocking" effect more pronounced for a larger rate  $a$. Consequently, less (or almost no) trajectories will reach
the default level $D$, creating stability of the system.

Next, we compare the loss distribution (distribution of number of defaults) for the coupled and independent cases. We
compute these loss distributions by Monte Carlo method using $10^4$ simulations, and with the same parameters as
previously.

\begin{figure}
\includegraphics[width=13cm,height=6cm]{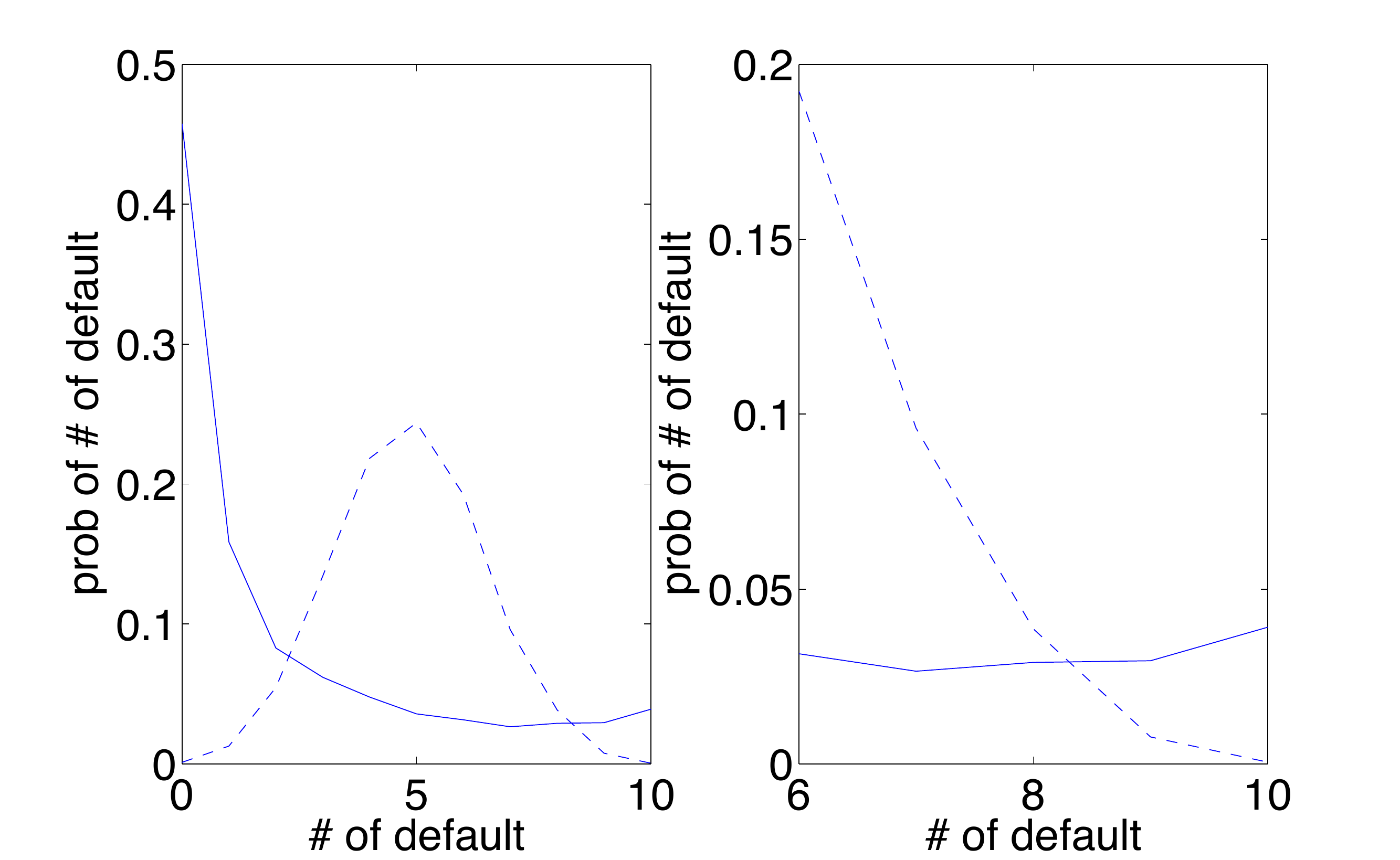}
\caption{On the left, we show plots of the loss distribution for the coupled diffusions with $a=10$ (solid line) and
for the independent Brownian motions (dashed line). The plots on the right show the corresponding tail
probabilities.}
\label{loss large alpha}
\end{figure}

\begin{figure}
\includegraphics[width=13cm,height=6cm]{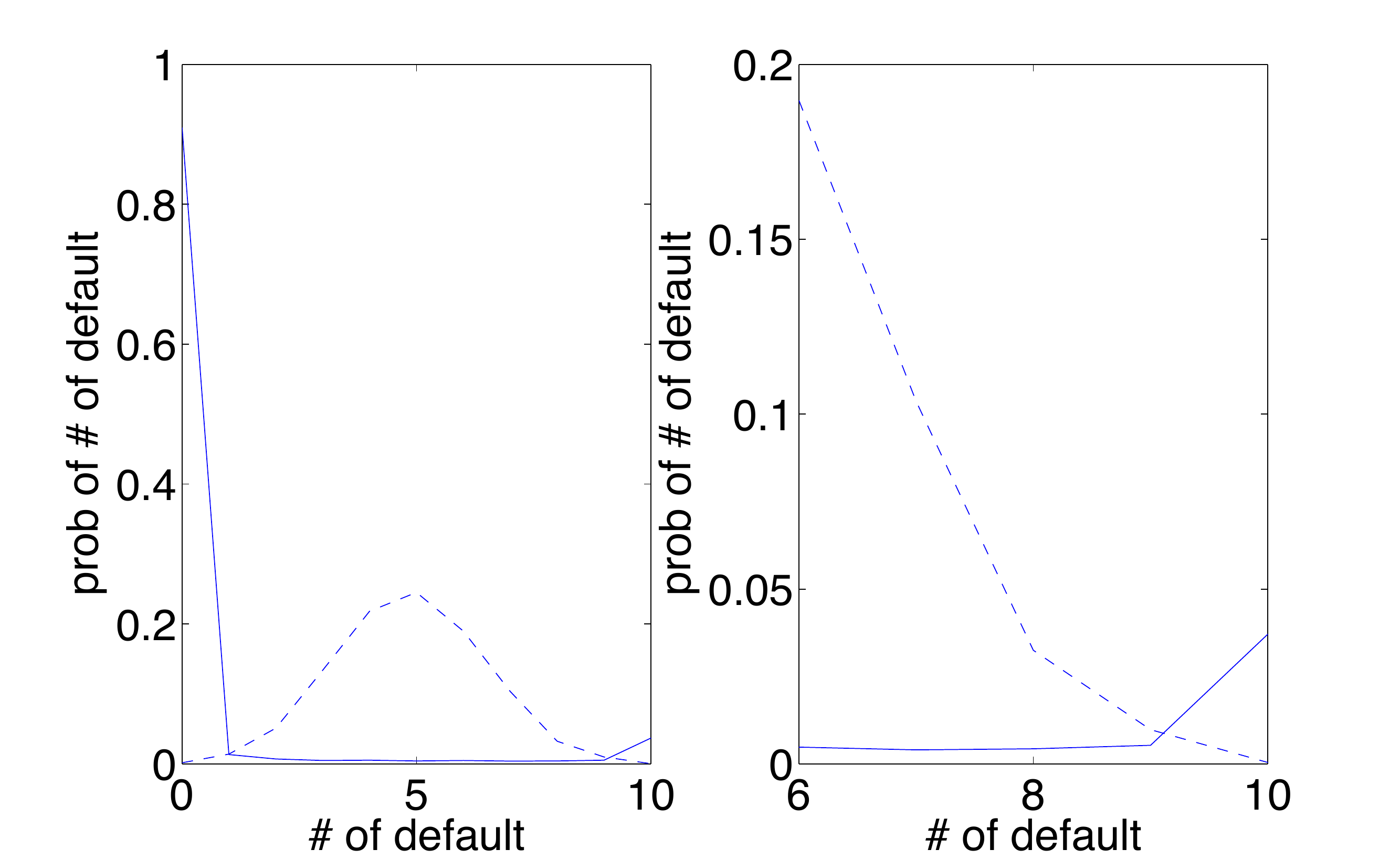}
\caption{On the left, we show plots of the loss distribution for the coupled diffusions with $a=100$ (solid line) and
for the independent Brownian motions (dashed line). The plots on the right show the corresponding tail
probabilities.}
\label{loss huge alpha}
\end{figure}

In the independent case ($a=0$), the loss distribution is Binomial($N,p$) with parameter $p$ given by
\ban
p=\PP\left(\min_{0\leq t\leq T}(\sigma W_t)\leq D\right)
= 2\Phi \left(\frac{D}{\sigma \sqrt{T}}\right),
\ean
where $\Phi$ denotes the ${\cal N}(0,1)$-cdf, and we  used the explicitly known distribution of the  minimum of a
Brownian motion.
With our choice of parameters, we have $p\approx0.5$ and therefore, the corresponding loss distribution is almost
symmetric as can be seen on the left panels (dashed lines) in
 Figures \ref{loss large alpha} and \ref{loss huge alpha}. 
We see that increasing $a$, that is the rate of inter-bank borrowing and lending, pushes most of the mass to zero
default, in other words, it improves the stability of the system by keeping the diffusions near zero (away from
default) most of the time. However, we also see that there is a small but non-negligible probability, that almost all
diffusions reach the default level. On the right panels of Figures \ref{loss large alpha} and \ref{loss huge alpha} we
zoom on this tail probability. In fact,  this tail corresponds to the small probability of the ensemble average
reaching the default level, and to almost all diffusions following this average due to ``flocking" for large $a$.

\subsection{Mean-field Limit}\label{sec:limit}

From (\ref{couplednogame}) we see that the processes $(X^{i}_t)$'s are ``OUs" mean-reverting to their {\it ensemble
average} $(\overline{X}_t)$.
Summing up the equations in (\ref{couplednogame}) and using $X^{i}_0=0,\,  i=1,\dots,N$ , one observes that this
ensemble average is given by
\be\label{ensemble}
\overline{X}_t =  \frac{\sigma}{N}\sum_{i=1}^{N}W_{t}^{i},
\en
which is distributed as a Brownian motion with diffusion coefficient $\sigma/\sqrt{N}$ independently of $a$.

In the limit $N\to \infty$, the strong law of large numbers gives
\[
\frac{1}{N}\sum_{j=1}^{N}W_{t}^{j}\rightarrow0\quad a.s.\,,
\]
and, more generally, the processes $X_{t}^{i}$ given by
\[
X_{t}^{i}=\frac{\sigma}{N}\sum_{j=1}^{N}W_{t}^{j}+\sigma e^{-a t}\int_0^t e^{a s}d W_s^{i}
-\frac{\sigma}{N}\sum_{j=1}^{N}\left(e^{-a t}\int_0^t e^{a s}d W_s^{j}\right)\,,
\]
converge to the independent
OU processes $\sigma e^{-\alpha t}\int_0^t e^{\alpha s}d W_s^{i}$ with long-run mean zero.
This is in fact a simple example of a mean-field limit and propagation of chaos studied in general in
\cite{Sznitman1991}.

\subsection{Large Deviations and Systemic Risk}\label{sec:ld}

In this section, we focus on the event where the ensemble average given by (\ref{ensemble}) reaches the default level.
The probability of this event is small (when $N$ becomes large), and is given by the theory of Large Deviations. In our
simple example, this probability can be computed explicitly as follows:
\ba
\PP\left(\min_{0\leq t\leq T}\left(\frac{\sigma}{N}\sum_{i=1}^{N}W_{t}^{i}\right)\leq {D}\right) &=&
\PP\left(\min_{0\leq t\leq T}\widetilde{W}_t \leq \frac{D\sqrt{N}}{\sigma}\right)\nonumber\\
&=&2\Phi\left(\frac{D\sqrt{N}}{\sigma\sqrt{T}}\right)\,,\label{probasystemic}
\ea
where $\widetilde{W}$ is a standard Brownian motion and $\Phi$ is the ${\cal N}(0,1)$-cdf. Therefore, using classical equivalent for the Gaussian cumulative
distribution function, we obtain
\be
\lim_{N\to\infty}-\frac{1}{N}\log\PP\left(\min_{0\leq t\leq
T}\left(\frac{\sigma}{N}\sum_{i=1}^{N}W_{t}^{i}\right)\leq {D}\right)=\frac{D^2}{2\sigma^2T}\,.
\en
In other words, for a large number of banks, the probability that the ensemble average reaches the default barrier is
of order $\exp(-D^2N/(2\sigma^2T))$. Recalling (\ref{ensemble}),
we identify
\be\label{systemic event}
\left\{\min_{0\leq t\leq T}\overline{X}_t \leq {D}\right\} 
\en
as a {\it systemic event}. Observe that this event does not depend on $a>0$, in other words, increasing stability by
increasing  the rate of borrowing and lending $a$ does not prevent a systemic event where a large number of banks
default. In fact, once in this event, increasing  $a$ creates even more defaults by ``flocking to default". This is
illustrated in the Figure \ref{loss huge alpha}, where $a=100$ and the probability of systemic risk is roughly $3\%$
(obtained using formula  (\ref{probasystemic})).

To summarize this section, our simple model with prescribed dynamics (no game) and independent noises shows that ``lending and borrowing
improves stability but also contributes to systemic risk". We have quantified this behavior and identified the crucial
role played by the inter-bank rate of borrowing and lending.

\subsection{Systemic Risk and Common Noise}\label{sec:commonnoise}

In this section, we discuss the coupled diffusions driven by correlated Brownian motions $(\widetilde{W}^{i}_t)$ and without control ($\alpha^i=0$). The dynamics (\ref{coupled}) or (\ref{coupled2}) becomes:
 
\be\label{correlated coupled}
dX_{t}^{i}=a\left(\frac{1}{N}\sum_{j=1}^{N}X_{t}^{j}-X_{t}^{i}\right)dt+\sigma\left(\rho
dW^0_{t}+\sqrt{1-\rho^{2}}dW_{t}^{i}\right),\quad i=1,\cdots,N,
\en
where $\left(W^0_t,W_t^i,i=1,\cdots, N\right)$ are independent standard Brownian motions and $W^0_{t}$ is a common noise. As previously, we calculate the ensemble average
\ban
\frac{1}{N}\sum_{i=1}^{N}X_{t}^{i}=\frac{\sigma}{N}\sum_{i=1}^{N}\widetilde{W}_{t}^{i}&=&\sigma\left(\rho
W^0_{t}+\frac{\sqrt{1-\rho^{2}}}{N}\sum_{i=1}^{N}W_{t}^{i}\right)\\
&\overset{{\cal D}}{=}&\sigma \sqrt{\rho^2+\frac{(1-\rho^{2})}{N}}B_{t},
\ean
where $B_{t}$ is a standard Brownian motion. Moreover, the explicit solution for $X_{t}^{i}$ is
\[
X_{t}^{i}=\sigma\rho
W^0_{t}+\sigma\sqrt{1-\rho^{2}}\left(\frac{1}{N}\sum_{j=1}^{N}W_{t}^{j}+\int_{0}^{t}e^{a(s-t)}dW_{s}^{i}-\frac{1}{N}\sum_{j=1}^{N}\int_{0}^{t}e^{a(s-t)}dW_{s}^{j}\right).
\]
The probability of the systemic event (\ref{systemic event}) becomes
\begin{eqnarray*}
\PP\left(\min_{0\leq s\leq T}\frac{1}{N}\sum_{i=1}^{N}X_{s}^{i}<D\right) & = & \PP\left(\min_{0\leq s\leq
T}B_{s}<\frac{D}{\sigma}\sqrt{\frac{N}{N\rho^{2}+(1-\rho^{2})}}\right)\\
 & = & 2\Phi\left(\frac{D}{\sigma\sqrt{T}}\sqrt{\frac{N}{N\rho^{2}+(1-\rho^{2})}}\right).
\end{eqnarray*}

\begin{figure}
\includegraphics[width=13cm,height=6cm]{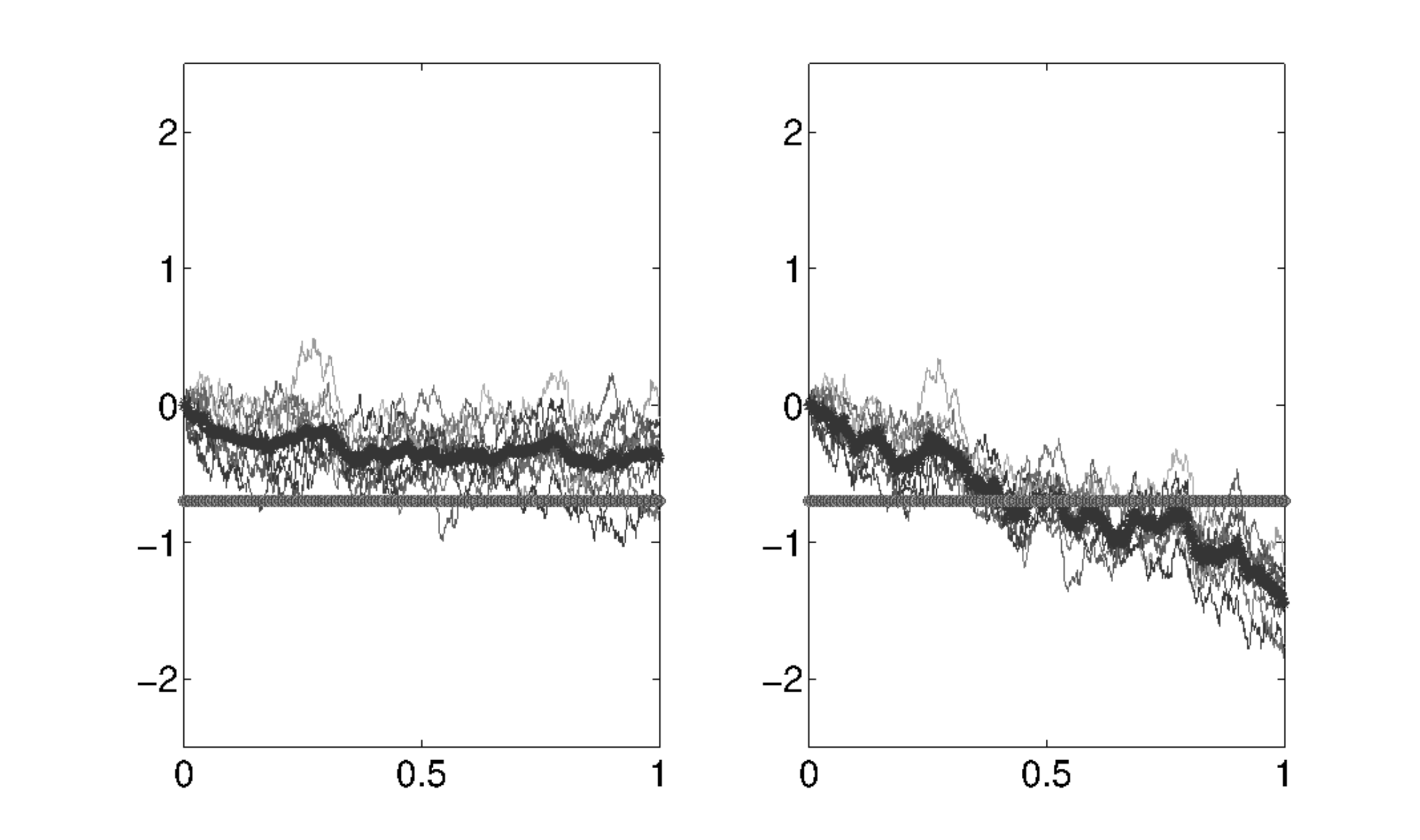}
\caption{One realization of $N=10$ trajectories of the coupled diffusions with independent Brownian motions
(\ref{couplednogame}) (left plot) and trajectories of the coupled diffusions with correlated Brownian motions with
$\rho=0.5$ (\ref{correlated coupled}) (right plot) using the common $a=10$. The solid horizontal line represents
the ``default" level $D=-0.7$.}
\label{large rho}
\end{figure}

From the  formula above, we see that in the correlated case ($\rho \ne 0$), the probability of systemic risk does not vanish as $N$ becomes large, instead it converges to $2\Phi\left(\frac{D}{\sigma |\rho|\sqrt{T}}\right)$. This is in dramatic contrast with the independent case ($\rho=0$) where the probability of systemic risk is exponentially small in $N$. We illustrate this instability created by the common noise in Figure \ref{large rho}.

\section{Constructions of Exact Nash Equilibria}\label{sec:game}

We now return to the model where each bank controls its rate of borrowing and lending and we search for Nash equilibria of the system  (\ref{coupled2}-\ref{fo:g}). We first construct open-loop equilibria using the Pontryagin stochastic maximum principle leading naturally to the solution of Forward-Backward SDEs (FBSDEs). Next, we construct closed-loop Markovian equilibria using two different approaches: the first one based on a modified version of Pontryagin stochastic maximum principle used in the open-loop case, and the other one based on the dynamic programming principle leading to the solution Hamilton-Jacobi-Bellman (HJB) PDEs. 
Recall that in all cases we try to find an equilibrium for the following borrowing-and-lending problem. The dynamics of the log-capitalizations $X^i_t$ for $i=1,\cdots,N$ are in the form:
\ba\label{Xit}
dX^i_t&=&\left[a(\overline{X}_t-X^i_t)+\alpha^i_t\right]dt +\sigma \bigg(\sqrt{1-\rho^2} dW^i_t+\rho dW^0_t\bigg),
\ea
where $W^i_t, i=0,1,\dots, N$  are independent  Brownian motions, $\sigma>0$ and $a\geq 0$.
Bank $i\in\{1,\cdots,N\}$ controls its rate of lending and borrowing (to a central bank) at time $t$ by choosing the control $\alpha^i_t$ in order to minimize
\ba\label{objectives}
J^i(\alpha^1,\cdots,\alpha^N)=\EE\left\{\int_0^T f_i(X_t,\alpha^i_t)dt+g_i(X_{T}^{i})\right\},
\ea
with
\ba
f_i(x,\alpha^i)&=&\left[\frac{1}{2}(\alpha^i)^2-q\alpha^i(\overline{x}-x^i)+\frac{\epsilon}{2}(\overline{x}-x^i)^2\right],\label{fi}\\
g_i(x)&=&\frac{c}{2}\left(\overline{x}-x^{i}\right)^2,\label{gi}
\ea
and where $f_i(x,\alpha)$ is convex in $(x,\alpha)$ under the assumption $q^2\leq \epsilon$.

Our model falls in the class of {\it Linear-Quadratic} (LQ) Mean-Field games.
What differentiates the various problems is the set of admissible strategies $\{\alpha^i_t, i=1,\cdots,N\}$ searched for equilibria.

\subsection{Open-Loop Equilibria}\label{sec:openloop}

In the deterministic case ($\sigma=0$), the open-loop problem corresponds to searching for an equilibrium among strategies which are (deterministic) functions  $\{\alpha^i_t, i=1,\cdots,N\}$ given at time $t=0$ and from which  $\{(X^i_t), i=1\cdots,N\}$ are deduced by (\ref{Xit}). See for instance \cite{basar}.

In the stochastic case ($\sigma>0$), the open-loop problem consists in searching for an equilibrium among strategies $\{\alpha^i_t, i=1,\cdots,N\}$ which are adapted processes satisfying some integrability property such as $\EE\left(\int_0^T|\alpha^i_t|dt\right)<\infty$, and most importantly, from which the dynamics of $\{(X^i_t), i=1\cdots,N\}$ are deduced by (\ref{Xit}). See \cite{Carmona542} for example.

Using the Pontryagin approach, the Hamiltonian for bank $i$ is given by
\begin{align}
\label{Hamiltonian}
&H^i(x^1,\cdots,x^N,y^{i,1},\cdots,y^{i,N},\alpha^1,\cdots,\alpha^N)\\
&=\sum_{k=1}^N\left[a(\o x-x^k)+\alpha^k\right]y^{i,k}
+\frac12(\alpha^i)^2-q\alpha^i(\o x-x^i)+\frac{\epsilon}{2}(\o x-x^i)^2.\nonumber
\end{align}
The forward dynamics of the states can be rewritten as:
\be\label{ol-dXi}
dX^i_t=\partial_{y^{i,i}}H^i(X_t,Y^{i}_t,\alpha_t)dt+\sigma \bigg(\sqrt{1-\rho^2} dW^i_t+\rho dW^0_t\bigg),
\en
with initial conditions $X^i_0=x^i$. Even though they do not appear (yet) in the above forward dynamics,
the processes $Y^i_t=(Y_t^{i,j}, j=1,\cdots,N)$ are the adjoint processes satisfying the backward equations
\be\label{ol-dYij}
dY_t^{i,j}=-\partial_{x^j}H^i(X_t,Y^{i}_t,\alpha_t)dt+\sum_{k=0}^NZ^{i,j,k}_tdW^k_t,
\en
with terminal conditions $Y^{i,j}_T=\partial_{x^j}g_i(X_T)$ and where the processes $Z^{i,j,k}_t$ are adapted and square integrable. The necessary condition of the Pontryagin stochastic maximum principle suggests that one minimizes the Hamiltonian $H^k$  with respect to $\alpha^k$. See for example the discussion of the Isaacs conditions in \cite{Carmona542}. This leads to the  choice:
\be\label{alpha-open-loop}
\hat\alpha^k=-y^{k,k}+q(\o x-x^k).
\end{equation}
In order to prove that these candidates actually form a Nash equilibrium, we assume that all  players  are making that choice, and let player $i$ finds his best response by solving the BSDE \eqref{ol-dYij} to identify his own adjoint process. The partial derivatives of the Hamiltonian $H^i$ are:
\ban
\partial_{y^{i,i}}H^i&=&(a+q)(\o x-x^i)-y^{i,i},\\
\partial_{x^j}H^i&=&\frac{a+q}{N}\sum_{k=1}^N (y^{i,k}-y^{i,j})+(\epsilon-q^2)(\o x-x^i)(\frac{1}{N}-\delta_{i,j})
\ean
where we used the usual notation $\delta_{i,j}=1$ if $i=j$ and $\delta_{i,j}=0$ if $i\ne j$. 

The particular forms of the Hamiltonian $H^i$ and the terminal conditions for the backward equations  suggest the ansatz
\be\label{ansatz-open-loop}
Y_t^{i,j}=\phi_t(\frac{1}{N}-\delta_{i,j})(\o X_t-X^i_t),
\en
where $\phi_t$ is a deterministic function satisfying the terminal condition $\phi_T=c$. 
Using this ansatz, a careful computation shows that the backward equations (\ref{ol-dYij}) become
\be\label{ol-dYij2}
dY_t^{i,j}=(\frac{1}{N}-\delta_{i,j})(\o X_t-X^i_t)\left[(a+q)\phi_t-(\epsilon-q^2)\right]dt+
\sum_{k=0}^NZ^{i,j,k}_tdW^k_t.
\en
The forward equation (\ref{ol-dXi}) becomes
\be\label{ol-dXi2}
dX^i_t=\left[a+q+(1-\frac{1}{N})\phi_t\right](\o X_t-X^i_t)dt+\sigma \bigg(\sqrt{1-\rho^2} dW^i_t+\rho dW^0_t\bigg),
\en
which by summation gives
\be\label{ol-dXbar}
d\o X_t=\sigma\rho dW^0_t+\sigma\sqrt{1-\rho^2}\left(\frac{1}{N}\sum_{k=1}^NdW^i_t\right).
\en
Consequently, one obtains
\be\label{dXbar-Xi}
d(\o X_t-X^i_t)=-\left[a+q+(1-\frac{1}{N})\phi_t\right](\o X_t-X^i_t)dt+\sigma\sqrt{1-\rho^2}\left(\frac{1}{N}\sum_{k=1}^NdW^k_t-dW^i_t\right).
\en
Differentiating the ansatz (\ref{ansatz-open-loop}) and using (\ref{dXbar-Xi}), we get
\ba\label{ol-dYij3}
dY_t^{i,j}&=&(\frac{1}{N}-\delta_{i,j})(\o X_t-X^i_t)\left[\dot\phi_t-\phi_t\left(a+q+(1-\frac{1}{N})\phi_t\right)\right]dt\nonumber\\
&&+\phi_t(\frac{1}{N}-\delta_{i,j})\sigma\sqrt{1-\rho^2}\left(\frac{1}{N}\sum_{k=1}^NdW^k_t-dW^i_t\right)
\ea
where $\dot\phi_t$ denotes the time-derivative of $\phi_t$.
Comparing the two It\^o decompositions  (\ref{ol-dYij2}) and (\ref{ol-dYij3}), the martingale terms give 
 the processes $Z^{i,j,k}_t$
 $$
Z^{i,j,0}_t=0,\quad Z^{i,j,k}_t=\phi_t\sigma\sqrt{1-\rho^2}(\frac{1}{N}-\delta_{i,j})(\frac{1}{N}-\delta_{i,k})\,\mbox{for }\,\, k=1,\cdots,N,
$$
 which turn out to be determistic in our case and hence adapted. Identifying the drift terms show that the function $\phi_t$ must satisfy the scalar Riccati equation
  \be\label{olRiccati}
 \dot\phi_t=2(a+q)\phi_t+(1-\frac{1}{N})\phi_t^2-(\epsilon-q^2),
 \en
with the terminal condition $\phi_T=c$. This equation can be solved explicitly. We defer the solution to Section \ref{sec:comparison} where we will provide a comparison with the closed-loop case for which a similar equation appears. We also defer to Section \ref{sec:financial} the discussion of the financial interpretation of this equilibrium which will be similar to the one for the  closed-loop equilibrium. 
 
 Note that the form (\ref{alpha-open-loop}) of the control $\alpha^i_t$,  and the ansatz (\ref{ansatz-open-loop}) combine to give:
 \be\label{ol-alphai}
 \alpha^i_t=\left[q+\phi_t(1-\frac{1}{N})\right](\o X_t-X^i_t).
 \en
It is interesting to remark that these controls, while constructed to form an open-loop equilibrium, are in fact in closed-loop feedback form! In this equilibrium, each bank $i$ can implement its strategy by knowing 
$ \o X_t-X^i_t$. Further implications will be discussed in Section \ref{sec:financial}, but we note here that since  the observation of $ \o X_t-X^i_t$ is needed, it is natural to search for equilibria when the admissible strategies  allow for the use of this observation, that is closed-loop strategies which are discussed next.

\subsection{Closed-Loop Equilibria, still via the FBSDE Approach}\label{sec:closedloop}

In this section we solve for an exact Nash equilibrium in closed-loop form when the players/banks at time $t$ have complete information of the states of all the other players at time $t$, or in other words we allow feedback strategies.

In this context, when all the other players have chosen strategies in feedback form given by deterministic functions $\alpha^k(t,x)$ of time and state, the Hamiltonian of player $i$ is given by (see \cite{Carmona542}):
\begin{align}
\label{fo:Hi}
&H^i(x,y^{i,1},\cdots,y^{i,N},\alpha^1(t,x),\cdots,\alpha^i_t,\cdots,\alpha^N(t,x))\\
=&\sum_{k\ne i}\left[a(\o x-x^k)+\alpha^k(t,x)\right]y^{i,k}
+\left[a(\o x-x^i)+\alpha^i\right]y^{i,i} \nonumber\\
&+\frac12(\alpha^i)^2-q\alpha^i(\o x-x^i)+\frac{\epsilon}{2}(\o x-x^i)^2,\nonumber
\end{align}
As in the open-loop case but now with the Hamiltonian (\ref{fo:Hi}), the 
forward dynamics of the states for $i=1,\cdots,N$ are given by
(\ref{ol-dXi}) and the backward equations are as in (\ref{ol-dYij}).
Minimizing $H^i$ over $\alpha^i$ gives the choices:
\begin{equation}
\label{fo:hatalpha}
\hat\alpha^i=-y^{i,i}+q(\o x-x^i),\qquad i=1,\cdots,N,
\end{equation}
and we again make the ansatz
\begin{equation}
\label{fo:ansatz}
Y^{i,j}_t=\eta_t\bigg(\frac1N-\delta_{i,j}\bigg)(\o X_t-X^i_t),
\end{equation}
where $\eta_t$ is a deterministic function satisfying the terminal condition $\eta_T=c$.
With  the choices \eqref{fo:hatalpha} we get 
\ban
\alpha^k(t,x)&=&\left[q+\eta_t(1-\frac{1}{N})\right](\o x-x^k),\\
\partial_{x^j}\alpha^k(t,x)&=&\left[q+\eta_t(1-\frac{1}{N})\right](\frac{1}{N}-\delta_{k,j}),
\ean
and a careful computation using the Hamiltonian (\ref{fo:Hi}) reduces the backward equations  to
\begin{eqnarray}
\label{fo:backward2}
dY^{i,j}_t&=&-\partial_{x^j}H^idt+\sum_{k=0}^NZ^{i,j,k}_tdW^k_t\nonumber\\
&=&(\frac{1}{N}-\delta_{i,j})(\o X_t-X_t^i)\bigg[(a+q)\eta_t-\frac{1}{N}(\frac{1}{N}-1)\eta_t^2+q^2-\epsilon
\bigg]dt\nonumber\\
&&+\sum_{k=0}^NZ^{i,j,k}_tdW^k_t,
\end{eqnarray}
with  terminal conditions $Y^{i,j}_T=c(\frac{1}{N}-\delta_{i,j})(\overline{X}_T-X^i_T)$.

The forward equations  become:
\ba
\label{Xicontrolled}
dX^i_t&=&\partial_{y^{i,i}}H^idt+\sigma \bigg(\sqrt{1-\rho^2} dW^i_t+\rho dW^0_t\bigg)\\
&=&\left[a+q+ (1-\frac{1}{N})\eta_t\right](\overline{X}_t-X^i_t)dt +\sigma \bigg(\sqrt{1-\rho^2} dW^i_t+\rho dW^0_t\bigg), \nonumber
\ea
and by summation  we deduce
\be\label{Xbarcontrolled}
d\o X_t=\sigma \bigg(\rho dW^0_t +\sqrt{1-\rho^2}\frac{1}{N}\sum_{k=1}^NdW^k_t \bigg).
\en
Differentiating the ansatz \eqref{fo:ansatz} and using the form (\ref{Xicontrolled})-(\ref{Xbarcontrolled}) of the forward dynamics, we get:
\begin{eqnarray}
\label{fo:backward3}
&&dY^{i,j}_t=\bigg(\frac1N-\delta_{i,j}\bigg)(\o X_t-X^i_t)\bigg[\dot \eta_t-\eta_t\left(a+q+(1-\frac{1}{N})\eta_t\right)\bigg]dt\nonumber\\
&&
\qquad\quad+\eta_t(\frac{1}{N}-\delta_{i,j})\sigma\sqrt{1-\rho^2}\sum_{k=1}^N(\frac{1}{N}-\delta_{i,k})dW^k_t.
\end{eqnarray}
 Next, identifying term by term the two It\^o decompositions \eqref{fo:backward2} and \eqref{fo:backward3}, we obtain from the martingale terms
$$
Z^{i,j,0}_t=0,\qquad Z^{i,j,k}_t=\eta_t\sigma\sqrt{1-\rho^2}(\frac{1}{N}-\delta_{i,j})(\frac{1}{N}-\delta_{i,k})\,\mbox{for }\,\, k=1,\cdots,N,
$$
which are indeed adapted and square integrable,
and from the drift terms:
$$
\dot \eta_t-\eta_t\left(a+q+(1-\frac{1}{N})\eta_t\right)=(a+q)\eta_t-\frac{1}{N}(\frac{1}{N}-1)\eta_t^2+q^2-\epsilon.
$$
Therefore, $\eta_t$ must satisfy the scalar Riccati equation 
\begin{equation}
\label{fo:riccati}
\dot \eta_t=2(a+q)\eta_t+(1-\frac{1}{N^2})\eta_t^2-(\epsilon-q^2),
\end{equation}
 with the terminal condition $\eta_T=c$.
Equation (\ref{fo:riccati}) admits the solution 
\be\label{eta-explicit}
\eta_{t}  = \frac{-(\epsilon-q^2)\left(e^{(\delta^+-\delta^-)(T-t)}-1\right)-c\left(\delta^+ e^{(\delta^+-\delta^-)(T-t)}-\delta^-\right)}
{\left(\delta^-e^{(\delta^+-\delta^-)(T-t)}-\delta^+\right)-c(1-\frac{1}{N^2})\left(e^{(\delta^+-\delta^-)(T-t)}-1\right)},
\en
where we used the notation
\be\label{deltas}
\delta^{\pm}=-(a+q)\pm\sqrt{R},
\en
with
\be\label{eq:R}
R:=(a+q)^{2}+\left(1-\frac{1}{N^2}\right)(\epsilon-q^{2})>0.
\en
Observe that $\eta_t$ is well defined for any $t\leq T$ since
 the denominator in (\ref{eta-explicit}) can be written as
\ban
&&-\left(e^{(\delta^+-\delta^-)(T-t)}+1\right)\sqrt{R}
-\left(a+q+c\left(1-\frac{1}{N^2}\right)\right)\left(e^{(\delta^+-\delta^-)(T-t)}-1\right),
\ean
which stays negative because $\delta^+-\delta^-=2\sqrt{R}>0$. In fact, using $q^2\leq \epsilon$, we see that $\eta_t$ is positive with $\eta_T=c$ as required as illustrated in Figure \ref{fig-eta}.

We delay the discussion of the  implications of our analysis in terms of banking system to Section \ref{sec:financial}. We first briefly present the dynamic programming approach to the problem which produces the same equilibrium as the one obtained in this section.

\subsection{Closed-Loop Equilibria via the HJB Approach}\label{sec:HJB}

In the Markovian setting, the value function of player $i$ is given by
\[
V^i(t,x)=\inf_\alpha\EE_{t,x}\left\{\int_t^T f_i(X_t,\alpha^i_t)dt+g_i(X_{T}^{i})\right\},
\]
with  the cost functions $f_i$ and $g_i$ given in (\ref{fi}) and (\ref{gi}), and where the dynamics of $X_t$ is given as before by: 
\[
dX^i_t=\left[a(\overline{X}_t-X^i_t)+\alpha^i_t\right]dt +\sigma \bigg(\sqrt{1-\rho^2} dW^i_t+\rho dW^0_t\bigg), \,\, i=1,\cdots,N.
\]
Using the dynamic programming principle in search for a closed-loop equilibrium, the corresponding HJB equations read
\ba\label{HJB}
\nonumber  \partial_{t}V^i&+&\inf_{\alpha}\bigg\{
\sum_{j=1}^N\left[a\left(\overline{x}-x^j\right)+{\alpha^j}\right]\partial_{x^j}V^i\\
&+&\frac{\sigma^2}{2}\sum_{j=1}^N\sum_{k=1}^N\left(\rho^2+\delta_{j,k}(1-\rho^2)\right)\partial_{x^jx^k}V^i\nonumber\\
&+&\frac{(\alpha^{i})^2}{2}-q\alpha^{i}\left(\overline{x}-x^i\right)+\frac{\epsilon}{2}(\o x-x^i)^2\bigg\} =0,
\ea
with terminal conditions $V^i(T,x)=\frac{c}{2}(\o x -x^i)^2$. Assuming that all controls $\alpha^j$ for $j\ne i$ are chosen, player $i$ will choose the optimal strategy $\hat\alpha^i=q(\o x-x^i)-\partial_{x^i}V^i$ where $V^i$ is still unknown.  Next, assuming that all players are following the strategies $\hat\alpha^i=q(\o x-x^i)-\partial_{x^i}V^i$, the HJB equations (\ref{HJB}) become
\ba\label{HJB2}
\nonumber  \partial_{t}V^i&+&
\sum_{j=1}^N\left[(a+q)\left(\overline{x}-x^j\right)-\partial_{x^j}V^j\right]\partial_{x^j}V^i\\
&+&\frac{\sigma^2}{2}\sum_{j=1}^N\sum_{k=1}^N\left(\rho^2+\delta_{j,k}(1-\rho^2)\right)\partial_{x^jx^k}V^i\nonumber\\
&+&\frac{1}{2}(\epsilon-q^2)\left(\overline{x}-x^i\right)^2+\frac{1}{2}(\partial_{x^i}V^i)^2
=0.
\ea
We then make the ansatz 
\ba\label{ansatzHJB}
V^i(t,x)=\frac{\tilde\eta_t}{2}(\o x -x^i)^2+\mu_t,
\ea
where $\tilde\eta_t$ and $\mu_t$ are deterministic functions satisfying $\tilde\eta_T=c$ and $\mu_T=0$ in order to match the terminal conditions for $V^i$. Note that the adjoint variables $y^{i,j}$ introduced in the FBSDE approach correspond to $\partial_{x^j}V^i$ and  the ansatz (\ref{ansatzHJB}) corresponds to the ansatz (\ref{fo:ansatz}).
The optimal strategies will be 
\ba\label{opt-alpha-HJB}
\hat\alpha^i=q(\o X_t-X^i_t)-\partial_{x^i}V^i=\left(q+(1-\frac{1}{N})\tilde\eta_t\right)(\o X_t-X^i_t),
\ea
and the controlled dynamics will become
\be\label{XicontrolledHJB}
dX^i_t=\left(a+q+ (1-\frac{1}{N})\tilde\eta_t\right)(\overline{X}_t-X^i_t)dt +\sigma \bigg(\sqrt{1-\rho^2} dW^i_t+\rho dW^0_t\bigg).
\en
Using
\[
\partial_{x^j}V^i=\tilde\eta_t(\frac{1}{N}-\delta_{i,j})\left(\overline{x}-x^i\right),\quad \partial_{x^jx^k}V^i=\tilde\eta_t(\frac{1}{N}-\delta_{i,j})(\frac{1}{N}-\delta_{i,k}),
\]
plugging into (\ref{HJB2}), and canceling terms in $\left(\overline{x}-x^i\right)^2$ and state-independent terms, we obtain
\ba
\dot{\tilde\eta}_t&=&2(a+q)\tilde\eta_t+(1-\frac{1}{N^2})\tilde\eta_t^2-(\epsilon-q^2),\label{eq:etatilde}\\
\dot \mu_t&=& -\frac{1}{2}\sigma^2(1-\rho^2)\left(1-\frac{1}{N}\right)\tilde\eta_t,\label{eq:mu}
\ea
with the terminal conditions $\tilde\eta_T=c$ and $\mu_T=0$.
Therefore, $\tilde\eta_t$ must satisfy the same Riccati equation  as the one  satisfied by $\eta_t$ (\ref{fo:riccati}). They have the same terminal conditions and by unicity of the solution for this equation, we deduce that $\tilde\eta_t=\eta_t$ for all  $t\leq T$ and, consequently, the two closed-loop approaches (FBSDE and HJB) produce the same equilibrium. The explicit solution for $\eta_t$ is given by (\ref{eta-explicit}) and, 
 furthermore, the solution $\mu_t$ of (\ref{eq:mu}) with the terminal condition $\mu_T=0$ is given by
\be\label{eq:mu-explicit}
\mu_t=  \frac{1}{2}\sigma^2(1-\rho^2)\left(1-\frac{1}{N}\right)\int_t^T\tilde\eta_s\,ds,
\en
and the value functions $V^i$ under this exact Nash equilibrium are given by (\ref{ansatzHJB}). Note that the correlation, quantified by the parameter $\rho$, affects the controls $\hat\alpha^i_t$ given by (\ref{opt-alpha-HJB}) only through the dynamics of $\o X_t- X^i_t$ since $\eta_t$ does not depend on $\rho$. However, it affects the value function $V^i$ given by (\ref{ansatzHJB}) also through the state-independent term $\mu_t$.

\subsection{Comparison of the Open- and Closed-Loop Equilibria}\label{sec:comparison}

Our analysis in Sections \ref{sec:openloop} and \ref{sec:closedloop} shows that the two equilibria we obtained are very similar. In fact, the only difference is that in the open-loop case we obtained the Riccati equation (\ref{olRiccati}) for the function $\phi_t$ (with a factor $(1-\frac{1}{N})$ in front of $\phi_t^2$), and in the closed-loop case we obtained the Riccati equation (\ref{fo:riccati}) for the function $\eta_t$ (with a factor $(1-\frac{1}{N^2})$ in front of $\eta_t^2$). 

In the closed-loop case we saw that the optimal strategy is given by
\[
 \alpha^i_t=\left[q+(1-\frac{1}{N})\eta_t\right](\o X_t-X^i_t),
 \]
 and the forward dynamics are:
 \[
dX^i_t=\left[a+q+ (1-\frac{1}{N})\eta_t\right](\overline{X}_t-X^i_t)dt +\sigma \bigg(\sqrt{1-\rho^2} dW^i_t+\rho dW^0_t\bigg), \nonumber
\]
with
\[
d\o X_t=\sigma \bigg(\rho dW^0_t +\sqrt{1-\rho^2}\frac{1}{N}\sum_{k=1}^NdW^k_t \bigg).
\]
These equations are identical in the open-loop case with $\eta_t$ replaced by $\phi_t$. Note that $\eta_t$ is given explicitly by formula (\ref{eta-explicit}) and $\phi_t$ can be obtained similarly by replacing the factor $(1-\frac{1}{N^2})$ by $(1-\frac{1}{N})$ (as well as in $\delta^{\pm}$ given by (\ref{deltas}) and $R$ given by (\ref{eq:R}) involved in that formula).

\begin{figure}
\includegraphics[width=13cm,height=6cm]{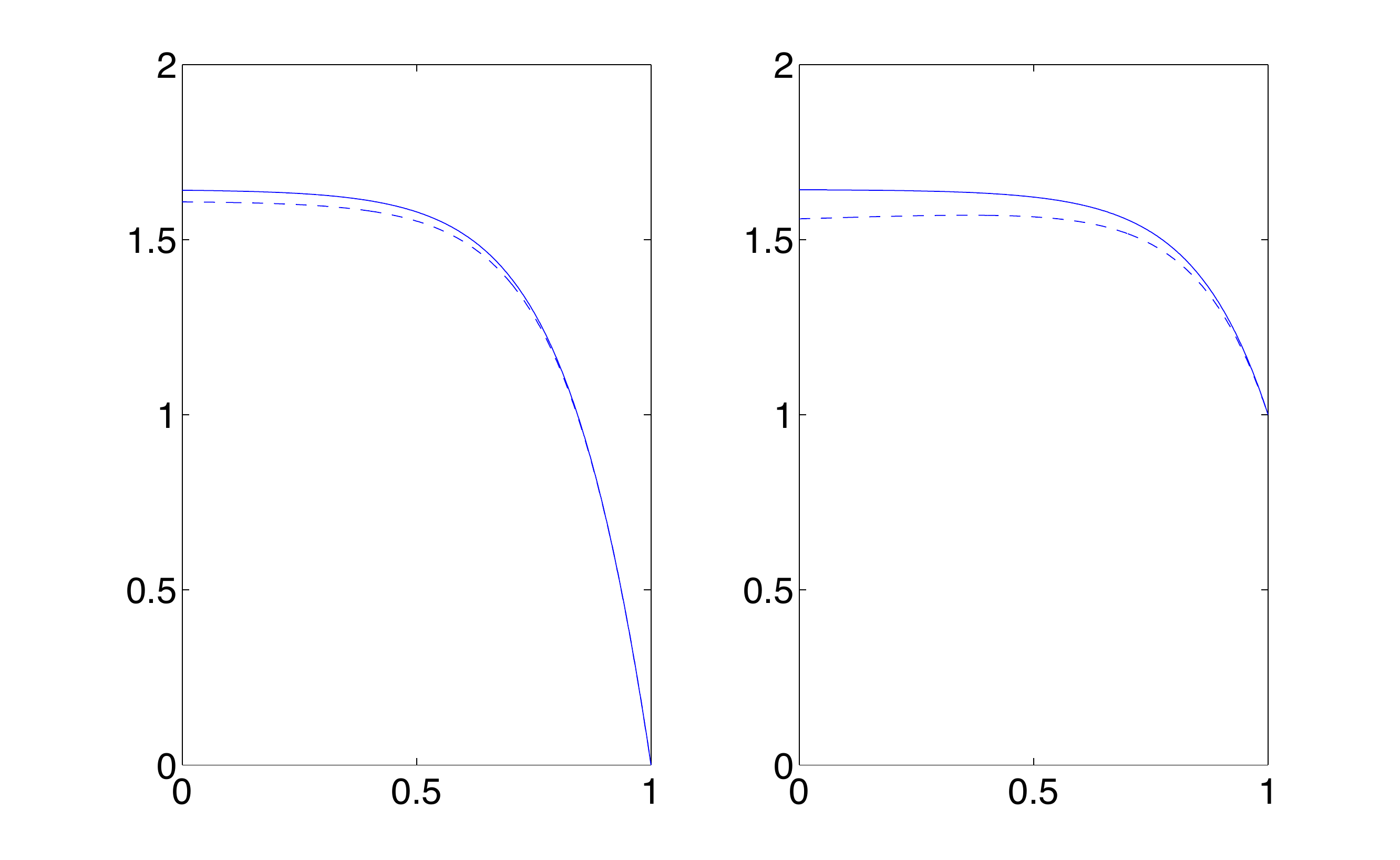}
\caption{Plots of $\phi_t$ (solid line) and $\eta_t$ (dashed line) with $N=10$, $a=1$, $q=1$, $\epsilon=10$, $T=1$, and $c=0$ on the left, $c=1$ on the right.}
\label{fig-eta-phi}
\end{figure}

In Figure \ref{fig-eta-phi}, we show the functions $\phi_t$ and $\eta_t$ involved respectively in the open-loop and closed-loop strategies. As expected, the difference is relatively small for $N=10$. However, it is enhanced by our choice of $\epsilon=10$ giving a rather large factor $\epsilon-q^2$ in front of $(1-\frac{1}{N})$ in the open-loop case or in front of $(1-\frac{1}{N^2})$ in the closed-loop case. Note that the presence of a terminal cost $c=1$ in the right panel produces a significant differen{c}e. 

The individual value functions can be calculated as follows. Considering for instance the closed-loop case, we want to calculate
\[
V^i(x)=\EE\left\{\int_0^T \left[\frac{1}{2}(\alpha^i)^2-q\alpha^i(\overline{X}_t-X_t^i)+\frac{\epsilon}{2}(\overline{X}_t-X_t^i)^2\right]dt+\frac{c}{2}(\overline{X}_T-X_T^i)^2\right\},
\]
 where $x$ is the initial position of the system and $(\alpha^i_t, X^i_t,\o X_t)$ are given by the equations above. Then, one easily obtains by direct computation
 \[
V^i(x)=\frac{1}{2} \int_0^T\left[\epsilon-q^2+(1-\frac{1}{N})^2\eta_t^2\right]\EE\left\{(\overline{X}_t-X_t^i)^2\right\}dt+\frac{c}{2}\EE\left\{(\overline{X}_T-X_T^i)^2\right\},
\]
with
\ban
\EE\left\{(\overline{X}_t-X_t^i)^2\right\}&=&(\o x-x^i)^2e^{-2\int_0^t(a+q+(1-\frac{1}{N})\eta_s)ds}\\
&&+(1-\frac{1}{N})\sigma^2(1-\rho^2)\int_0^t e^{-2\int_s^t(a+q+(1-\frac{1}{N})\eta_u)du}ds.
\ean
The formula in the open-loop case is simply obtained by replacing $\eta_t$ by $\phi_t$.

\begin{figure}
\includegraphics[width=13cm,height=7cm]{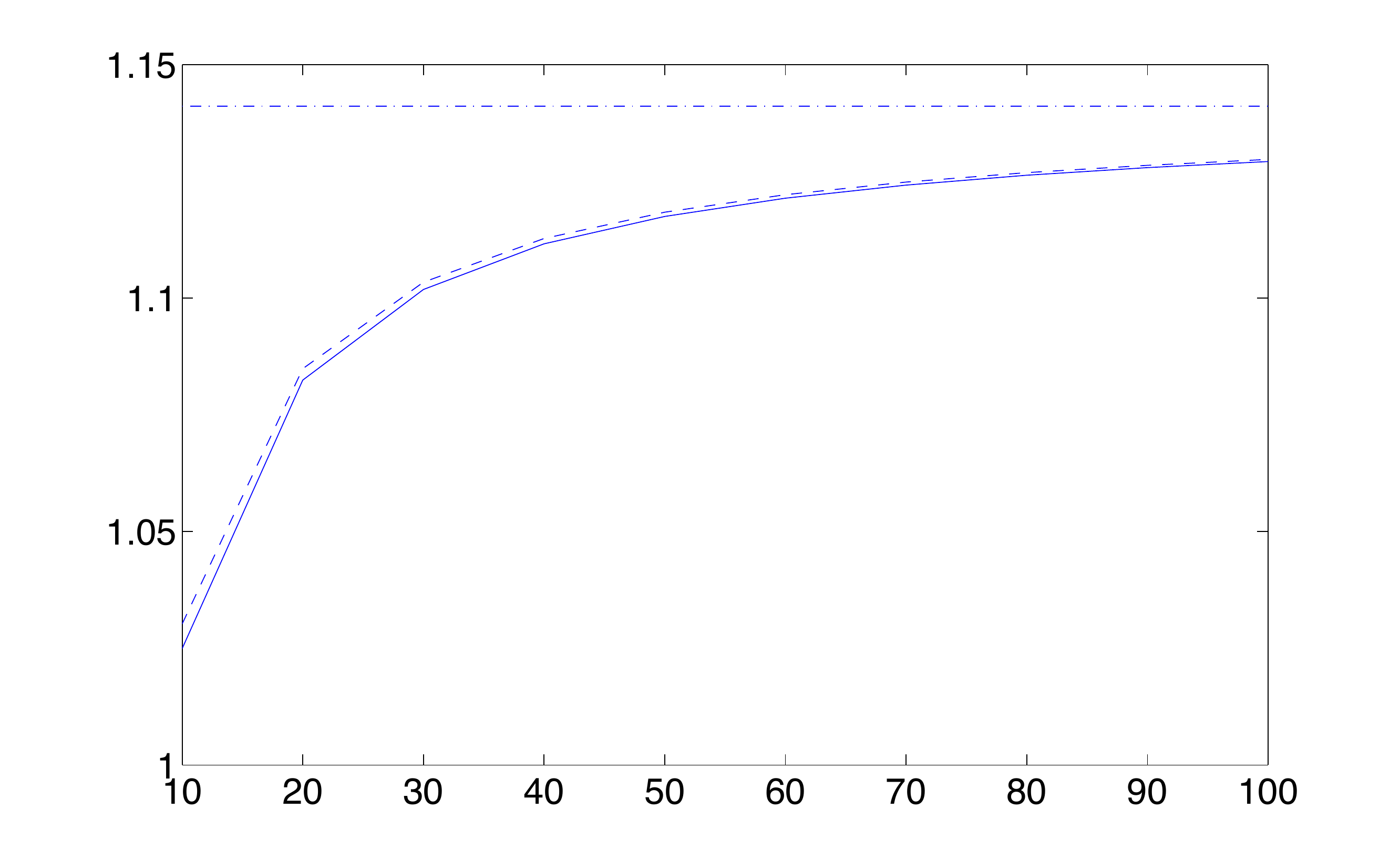}
\caption{Plots of the value function $V^i$ as $N$ increases: open loop (solid line), closed loop  (dashed line), and common limit as $N\to\infty$ (dotted line) with $a=1$, $q=1$,
$\epsilon=10$, $\rho=0.2$, $T=1$, and $c=10$.}
\label{fig-Vi}
\end{figure}

In Figure \ref{fig-Vi}, we compare the value functions $V^i$ in the open-loop and closed-loop equilibria for a choice of parameters and as $N\to \infty$.

\section{Financial Implications}\label{sec:financial}

The comments below made in the case of the closed-loop equilibrium with the function $\eta_t$ would be identical in the case of the open-loop equilibrium with $\eta_t$ replaced by $\phi_t$.

\begin{figure}
\includegraphics[width=13cm,height=6cm]{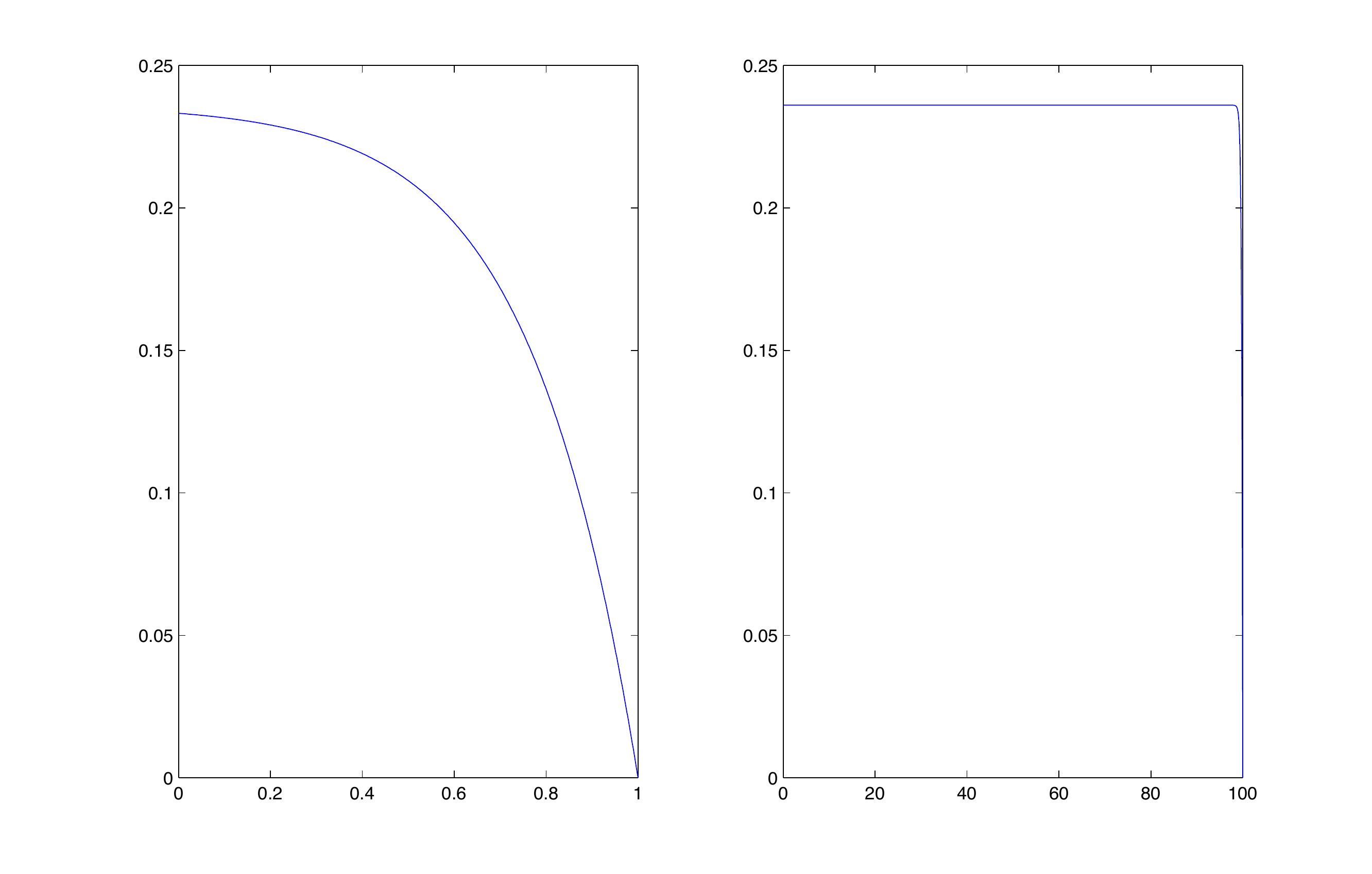}
\caption{Plots of $\eta_t$ with $c=0$, $a=1$, $q=1$, $\epsilon=2$ and $T=1$ on the left, $T=100$ on the right with $\o \eta\sim0.24$ (here we used $1/N\equiv 0$).}
\label{fig-eta}
\end{figure}

\begin{figure}
\includegraphics[width=13cm,height=7cm]{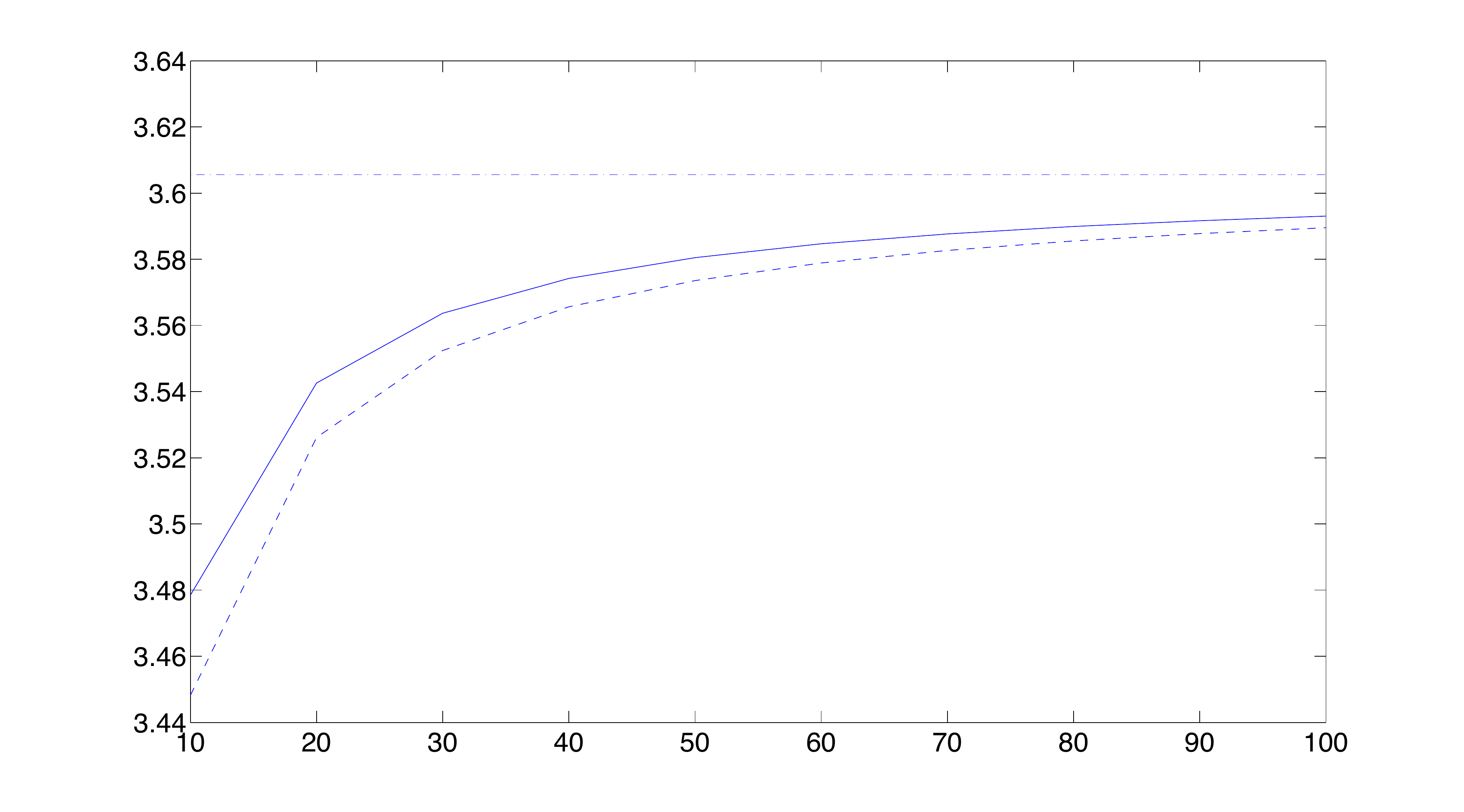}
\caption{Plots of the effective rate $A$ (\ref{eq:A}) for the open-loop equilibrium (solid line) and for the closed-loop equilibrium  (dashed line) with $a=1$, $q=1$, $\epsilon=10$, $T=1$,  and as $N$ increases. The dotted line shows the common limit  as $N\to \infty$}
\label{fig-A}
\end{figure}

\begin{enumerate}
\item
Once the function $\eta_t$ has been obtained in (\ref{eta-explicit}), bank $i$ implements its strategy by using its control $\hat\alpha^i$ given by (\ref{opt-alpha-HJB}). It requires its own log-reserve $X^i_t$ but also the average reserve $\o X_t$ which may or may not be known to the individual bank $i$. 
Observe that  the average $\o X_t$ is given by (\ref{Xbarcontrolled}), and 
 is identical to the average found in Section \ref{sec:commonnoise}. Therefore, systemic risk occurs in the same manner as in the case of  uncontrolled dynamics with or without common noise as presented respectively in Sections \ref{sec:commonnoise} and \ref{sec:ld}.  
\item
However, (\ref{XicontrolledHJB}) shows that the control affects the rate of borrowing and lending  by adding the time-varying component $q+ (1-\frac{1}{N})\eta_t$ to the uncontrolled rate $a$.
\item
In fact, from (\ref{XicontrolledHJB}) rewritten as 
\ba\label{Xicontrolled2}
dX^i_t&=&\left(a+q+ (1-\frac{1}{N})\eta_t\right)\frac{1}{N}\sum_{j=1}^N({X}^j_t-X^i_t)dt\\
 &&+\sigma \bigg(\sqrt{1-\rho^2} dW^i_t+\rho dW^0_t\bigg), \nonumber
\ea
we see that the effect of the banks using their optimal strategies corresponds to inter-bank borrowing and lending at the increased {\bf  effective rate}
\[
A_t:=a+q+ (1-\frac{1}{N})\eta_t
\]
 with no central bank (or a central bank acting as an instantaneous {\bf clearing house}). As a consequence, under this equilibrium, the system is operating as if  banks were borrowing from and lending to each other at the rate $A_t$, and the net effect is {\bf additional liquidity} quantified by the rate of lending/borrowing. 
 
 Note that the comment above is valid not only if $a>0$, in which case the effect of the game is to increase the rate of interbank lending and borrowing, but also if $a=0$, in which case the effect of the game is to ``create" an interbank lending and borrowing activity. In both cases, the central bank acts as a clearing house but needs to provide the information $\o X_t$ so that individual banks can implement their strategies.  

 \item
Observe that the presence of a common noise (quantified by $\rho$) does not affect the form of the optimal strategies (the function $\eta_t$ does not depend on $\rho$). However it affects the value function $V^i(t,x)$ and the dynamics $X^i_t$, and, as we have seen in   Section \ref{sec:commonnoise}, it has a drastic effect on systemic risk.

\item
It is also interesting to note that for $T$ large, most of the time ($T-t$ large), $\eta_t$ is mainly constant. For instance, with $c=0$, 
\[
\lim_{T\to\infty}\eta_t=\frac{\epsilon-q^2}{-\delta^-}:=\o \eta,
\]
as illustrated on right panel of Figure \ref{fig-eta}.
Therefore, in this infinite-horizon equilibrium, banks are borrowing and lending to each other at the constant rate 
\be\label{eq:A}
A:= a+q+(1-\frac{1}{N})\o\eta.
\en 

In Figure \ref{fig-A} we show the constant effective rates $A$ (for infinite horizon) for the open-loop and closed-loop equilibria as $N$ increases. Note that liquidity (quantified by the effective rate of lending/borrowing $A$) is higher under the open-loop equilibrium and increases with $N$.

\end{enumerate}

In the following section, we discuss asymptotics as $N\to\infty$.

\section{Approximate Nash Equilibria}
\label{sec:approximate}

Why would we want an approximate Nash equilibrium when we can compute an exact one?

The game presented in the previous section was essentially designed to provide explicit Nash equilibria, and the fact that  it does is already remarkable! However, slight modifications, even minor, dramatically change the equilibrium structures. For instance, the presence in the dynamics of the $X^i_t$ or in the objective functions $J^i$ of a nonlinear term in $\o X_t$ renders the computation of exact Nash equilibria hopeless. The  Mean Field Game strategy is based on 
the solution of effective equations in the limit $N\to\infty$, providing through the theory of the propagation of chaos, approximate Nash equilibria for the case $N$ finite and large. We review this strategy in the context of the model presented in the previous section, and compare its output to the exact solutions derived earlier.

\subsection{{The Mean Field Game / FBSDE Approach}}

If it was not for the presence of a common noise, we could apply the results of \cite{CarmonaDelarue_sicon} to obtain approximate Nash equilibria from the solution of the Mean Field Game (MFG).
Notice that linear quadratic MFGs are studied in \cite{Bensoussan_et_al} and \cite{CarmonaDelarueLachapelle}. While the latter does not include the cross term $-q\alpha(\o x-x^i)$ in the running cost of player $i$, the proofs of \cite{CarmonaDelarueLachapelle} apply \emph{mutatis mutandis} to the model of the present paper when $\rho=0$.
We review the MFG strategy. It is based on the following three steps:
\begin{enumerate}
\item
Fix $(m_t)_{t\geq0}$, which should be thought  of as a candidate for the  limit of $\overline{X}_t$ as $N\to\infty$: 
\[
m_t=\lim_{N\to\infty} \o X_t.
\]
Because of the presence of the common noise, $(m_t)_{t\geq0}$ is a process adapted to the filtration generated by $W^0$ and one should think of $m_t$ as a function of $(W^0_s)_{s\leq t}$;  
\item
Solve the one-player standard control problem
{\small$$
\inf_{\alpha=(\alpha_t)\in {\cal
A}}\EE\left\{\int_0^T\left[\frac{\alpha_t^2}{2}-q\alpha_t(m_t-X_t)+\frac{\epsilon}{2}(m_t-X_t)^2\right]dt+\frac{c}{2}(m_T-X_T)^2\right\},
$$}
subject to the dynamics
$$
dX_t=\left[a(m_t-X_t)+\alpha_t\right]dt +\sigma \left(\rho dW^0_t+\sqrt{1-\rho^2}dW_t\right),
$$
where $W^0_t$ and $W_t$ are independent Brownian motions, independent of the initial value $X_0$ which may be a square integrable random variable $\xi$.

\item
Solve the  fixed point problem: 
find $m_t$ so that $m_t=\EE [X_t\mid (W^0_s)_{s\leq t} ]$ for all $t$.
\end{enumerate}

We treat the above stochastic control problem as a problem of control of non-Markovian dynamics with random coefficients. The Hamiltonian of the system is given by
\ban
H(t,x,y,\alpha)=\left[a(m_t-x)+\alpha\right]y+\frac{1}{2}\alpha^2-q\alpha(m_t-x)+\frac{\epsilon}{2}(m_t-x)^2,
\ean
which is  strictly convex in $(x,\alpha)$ under the condition $q^2\leq \epsilon$ and attains its minimum at
\ban
\frac{\partial H}{\partial \alpha}=0\longrightarrow \hat{\alpha}=q(m_t-x)-y.
\ean
The corresponding {\it adjoint forward-backward equations} are given by
\ba
dX_t&=&\frac{\partial H}{\partial y}(\hat{\alpha}) dt+\sigma\left(\rho dW^0_t+\sqrt{1-\rho^2}dW_t\right)\label{forward}\\
&=&\left[(a+q)(m_t-X_t)-Y_t\right]dt+\sigma \left(\rho dW^0_t+\sqrt{1-\rho^2}dW_t\right),\quad X_0=\xi\nonumber\\
dY_t&=&-\frac{\partial H}{\partial x}(\hat{\alpha}) dt+ Z^0_tdW^0_t+Z_tdW_t\label{backward}\\
&=&\left[(a+q)Y_t+(\epsilon-q^2)(m_t-X_t)\right]dt\nonumber
+Z^0_tdW^0_t+Z_tdW_t,\\
&&\hskip 7cm Y_T=c(X_T-m_T),\nonumber
\ea
for some adapted square integrable processes $(Z^0_t,Z_t)$.
Despite its simple looking structure, such linear systems do not 
always have solutions. The existence of a solution in the present situation is argued in \cite{CarmonaDelarueLachapelle}
where a solution is shown to exist. To identify it in the present situation, we use the notation $m^X_t=\EE[X_t\mid  (W^0_s)_{s\leq t}]$ and
$m^Y_t=\EE[Y_t\mid  (W^0_s)_{s\leq t}]$. Taking conditional expectation given  $(W^0_s)_{s\leq t}$ in the second equation (\ref{backward}), and using the fact that in equilibrium 
(i.e. after solving for the fixed point), we have $m_t=m^X_t$ for all $t\leq T$ which in turn implies  $m^Y_T=c(m^X_T-m_T)=0$, we obtain: 
\be\label{eq:mYt}
m^Y_t=
-\int_t^Te^{(a+q)(s-t)}Z^0_sdW^0_s.
\en
Next, taking conditional expectations of both sides of equation \eqref{forward},
we deduce that, in equilibrium,  we shall have 
\be\label{eq:mXt}
dm^X_t=-m^Y_tdt+\rho\sigma dW^0_t.
\en
Now, we make the (educated) ansatz
\be\label{ansatzMFG-FBSDE}
Y_t=-\eta_t(m_t-X_t),
\en
for some deterministic function $t\hookrightarrow \eta_t$ to be determined. Differentiating this ansatz and using (\ref{forward}) and (\ref{eq:mXt}) leads to
\ba\label{Ito1}
dY_t&=&-\dot \eta_t(m_t-X_t)dt-\eta_td(m_t-X_t)\\
&=&\left[\left(-\dot \eta_t +\eta_t(a+q+\eta_t)\right)(m_t-X_t)+\eta_t m^Y_t\right]dt+\eta_t\sigma\sqrt{1-\rho^2}dW_t.\nonumber
\ea
Plugging the ansatz (\ref{ansatzMFG-FBSDE}) in (\ref{backward}) gives
\be\label{Ito2}
dY_t=\left[-(a+q)\eta_t+(\epsilon-q^2)\right](m_t-X_t)dt +Z^0_tdW^0_t+Z_tdW_t.
\en
Identifying the two It\^o decompositions (\ref{Ito1}) and (\ref{Ito2}), we deduce first from the martingale terms  that
$Z^0_t\equiv 0$ and $Z_t=\eta_t\sigma\sqrt{1-\rho^2}$. Thus, from (\ref{eq:mYt}) we obtain $m^Y_t=0$, and equating the drift terms we see that $\eta_t$ must be solution to the Riccati equation
\be\label{Riccati 38}
\dot \eta_t=2(a+q)\eta_t+\eta_t^2-(\epsilon-q^2),
\en
with terminal condition $\eta_T=c$. As expected, the solution for $\eta_t$ is given explicitly in (\ref{eta-explicit}) after taking the limit  $N\to \infty$.
Observe that from $m^Y_t=0$ and (\ref{eq:mXt}) we deduce that $m^X_t=\EE(\xi)+\rho\sigma W^0_t$ which will enter in the optimal control $(q+\eta_t)(m^X_t-X_t)$.

An important result of the theory of MFGs (see for example \cite{CarmonaDelarue_sicon}) is the fact that, once a solution to the MFG is found, on can use it to construct approximate Nash equilibria for the finitely many players games. Here, if one assumes that each player is given the information $\o X_t$, and if player $i$ uses the strategy
\[
\alpha^i_t=(q+\eta_t)(\o X_t-X^i_t),
\]
which is the limit as $N\to\infty$ of the strategy used in the finite players game, one sees how solving the limiting MFG problem can provide approximate Nash equilibria for which the financial implications are identical as the ones given in Section  \ref{sec:financial} for the exact Nash equilibria.


\subsection{{The Mean Field Game / HJB Approach}}

It is interesting to go through the derivation of the MFG solution using the HJB approach since it involves additional difficulties due to the presence of the common noise. In our toy model, it can be handled explicitly and we briefly outline this derivation.

\vskip 2pt
For Markovian strategies of the form $\alpha(t,x)$, the dynamics are given as previously by
\[
dX_t=\left[a(m_t-X_t)+\alpha(t,X_t)\right]dt +\sigma \left(\rho dW^0_t+\sqrt{1-\rho^2}dW_t\right).
\]
Here, we should think as $(W^0_t)$ as given, so that the forward equation for the conditional density of $X_t$ becomes a stochastic PDE (SPDE) which can be written as 
\[
dp_t=\left\{-\partial_x\left[\left(a(m_t-x)+\alpha(t,x)\right)p_t\right]+\frac{1}{2}\sigma^2(1-\rho^2)\partial_{xx}p_t\right\}dt-\rho\sigma(\partial_x p_t)dW^0_t,
\]
with the initial density $p_0$ being the density of $\xi$.  Here $\alpha(t,x)$ is given and  $m_t=\int xp_t(x)dx$.
Consequently, $m_t$, the conditional mean of $X_t$ given $W^0$  is a stochastic process which will turn out to be Markovian with its infinitesimal generator denoted by ${\cal L}^mdt+\rho\sigma (\partial_m) dW^0_t$. The HJB equation for the value function $V(t,x,m)$ can be written as 
\ban
dV&+& \left[\frac{1}{2}\sigma^2(1-\rho^2)\partial_{xx}V+{\cal L}^mV+(\partial_{xm}V)\frac{d\langle m,X\rangle}{dt}\right]dt\\
&+&\inf_\alpha\left\{\left[a(m-x)+\alpha\right]\partial_xV +\frac{\alpha^2}{2}-q\alpha(m-x)+\frac{\epsilon}{2}(m-x)^2\right\}dt\\
&+&\rho\sigma (\partial_mV) dW^0_t+ \rho\sigma(\partial_x V)dW^0_t=0.
\ean
Next, we minimize in $\alpha$ to get $\hat{\alpha}=q(m-x)-\partial_x V$, and we make the ansatz $V(t,x,m)=\frac{\eta_t}{2}(m-x)^2+\mu_t$. Plugging in the forward equation for $p_t$, multiplying by $x$,  and integrating with respect to $x$ gives
\[
dm_t=\rho\sigma dW^0_t.
\] 
 Therefore, conditionally in $W^0$, ${\cal L}^mV=0$ and $d\langle m,X\rangle=0$. Then, verifying that the ansatz satisfies the HJB equation, by canceling terms in $(m-x)^2$ we obtain that $\eta_t$ must satisfy the Riccati equation (\ref{Riccati 38}), and canceling state-independent terms leads to $\dot \mu_t=-\frac{1}{2} \sigma^2(1-\rho^2)\eta_t$ and therefore
 \[
 \mu_t=\frac{1}{2}\sigma^2(1-\rho^2)\int_t^T\eta_sds.
 \]
 
\subsection{{Controlling McKean-Vlasov Dyamics: Possibly a Different Solution}}

In this subsection we search for a different type of approximate equilibrium. As before, we require that all the banks use the same feedback function $\alpha$
to compute their lending and borrowing rate from their private information, but we now work with a different notion of equilibrium. We now base the notion of a critical point for the optimization on a simultaneous deviation of all the strategies at once.  So instead of stress-testing by perturbing the lending and borrowing policy one at a time like in the search for Nash equilibria, we perturb them simultaneously. This form of equilibrium was tentatively called \emph{franchised} equilibrium in \cite{CarmonaDelarue_ap}. See also
\cite{CarmonaDelarueLachapelle} for a comparison with the results of the MFG models. Our interest in this model comes from the results of these two papers which show that, despite strong similarities, these two models can lead to different solutions, even in the case of linear - quadratic models.
\vskip 4pt
The theory of the propagation of chaos implies that, when $N\to\infty$,
the individual states $X^i_t$ become independent and identically distributed, their common distribution being the law of the solution
of an equation of the McKean-Vlasov type:
\begin{equation}
\label{fo:mkv}
dX_t=[a(\EE[X_t]-X_t) +\alpha (t,X_t)]dt+\sigma dW_t,
\end{equation}
for some Wiener process $W$. See for example \cite{Sznitman1991} or \cite{JourdainMeleardWoyczynski}.
Accordingly, the expected cost that the players try to minimize becomes
$$
J(\phi)=\EE\bigg[\int_0^T f(X_t,\PP_{X_t},\alpha(X_t))dt+g(X_T,\PP_{X_T})\bigg],
$$
where we use the notation $\PP_Y$ to denote the law of the random variable $Y$. In the linear-quadratic case at hand, both functions $f$ and $g$ do not depend upon the entire distribution $\PP_{X_t}$, but only upon the first two moments $\EE[X_t]$ and $\EE[X^2_t]$.
The search for equilibrium considered in this subsection amounts to minimizing $J(\alpha)$ over a set of admissible feedback functions $\alpha$ under the constraint \eqref{fo:mkv}. The probabilistic approach to the solution of this problem is developed in \cite{CarmonaDelarue_ap} where a Pontryagin stochastic maximum principle is proved in full generality. Adjoint equations are identified and it is shown that there are different than in the MFG case. However, in the present situation, the fact that the cost functions $f(x,\mu,\alpha)$ and $g(x,\mu)$ depend only upon the distance from the argument $x$ to the mean of the measure $\mu$ implies that the expectation $\EE[Y_t]$ of the adjoint process vanishes, and the correction to the driver of the MFG adjoint equation vanishes as well. In the notation of \cite{CarmonaDelarueLachapelle}, this corresponds to the fact that $\o q=-q$ and $\o m_t=-m_t$, implying that the differences exhibited there between the solutions of linear quadratic MFG and McKean-Vlasov controlled dynamics do not exist in this instance. Again, this is due to the special form of interaction between the players through $\o X_t-X^i_t$.

\vskip 2pt
In any case, the two forms of approximate equilibria, though different in general, do coincide in our model of lending and borrowing!




\bibliographystyle{plain}

\bibliography{references}

\end{document}